\documentstyle[aps,pra,epsfig,twocolumn]{revtex}

\def\be{\begin{equation}}
\def\ee{\end{equation}}
\def\bea{\begin{eqnarray}}
\def\eea{\end{eqnarray}}
\def\bma{\begin{mathletters}}
\def\ema{\end{mathletters}}

\newcommand{\eins}{\mbox{$1 \hspace{-1.0mm}  {\bf l}$}}
\def\C{\hbox{$\mit I$\kern-.7em$\mit C$}}

\begin{document}
\draft

\title{Non--local Operations: Purification, storage, compression, tomography,
and probabilistic implementation}

\author{W. D\"ur and J. I. Cirac}

\address
{Institut f\"ur Theoretische Physik, Universit\"at Innsbruck,A-6020 Innsbruck, Austria}

\date{\today}

\maketitle

\begin{abstract}
We provide several applications of a previously introduced
isomorphism between physical operations acting on two systems
and entangled states \cite{Ci00}. We show: (i) how to implement
(weakly) non--local two qubit unitary operations with a small
amount of entanglement; (ii) that a known, noisy, non--local
unitary operation as well as an unknown, noisy, local unitary
operation can be purified; (iii) how to perform the tomography
of arbitrary, unknown, non--local operations; (iv) that a set of
local unitary operations as well as a set of non--local unitary
operations can be stored and compressed; (v) how to implement
probabilistically two--qubit gates for photons. We also show how
to compress a set of bipartite entangled states locally, as well
as how to implement certain non--local measurements using a
small amount of entanglement. Finally, we generalize some of our
results to multiparty systems.
\end{abstract}


\narrowtext

%-------------------------------------------------------------------------
%-------------------------------------------------------------------------
\section{Introduction}
%-------------------------------------------------------------------------
%-------------------------------------------------------------------------

In recent years, much of the theoretical effort in quantum
information (QI) theory was focused on establishing properties
of states and techniques to manipulate them. One of the main
purposes was ---and is--- the characterization and
quantification of entanglement properties of multiparticle
states, as entangled states play an important role in several
applications of QI. Many schemes and applications which involve
the manipulation of quantum states were discovered. Among them,
we have teleportation \cite{Be93T}, purification of noisy
entanglement \cite{Be93,Be96,De96}, quantum data compression
\cite{Sc97}, quantum cloning \cite{cloning}, and quantum
tomography \cite{vo89}.

In practice, entangled states are created by some physical
action. This suggest that establishing properties of operations
may play an important role in QI as well. First steps in this
direction have been recently reported
\cite{Za00,Du00H,Kr00,Za00I}. In particular, in Ref.
\cite{Du00H} the {\em entanglement capability} of interaction
Hamiltonians between two systems has been introduced and
analyzed. This quantity measures the maximum rate at which
entanglement can be produced given some particular interaction.
On the other hand, the entanglement cost for the implementation
of non--local operations was also considered recently
\cite{Go98,Ch00,Ei00,Co00}; in particular, several examples, all
dealing with an integer number of ebits required for the
implementation of certain non--local operations, were
introduced.

In Ref. \cite{Ci00} we introduced an isomorphism which relates
physical operations (completely positive maps (CPM) ${\cal E}$)
and states (positive operators $E$). This isomorphism turns out
to be an important ingredient in the understanding of
entanglement properties of operations in general. In this paper,
we will first review the results obtained in Ref. \cite{Ci00}:

\begin{itemize}
\item In order to study the separability and entangling properties of
operations ${\cal E}$, it suffices to study the separability
properties of the associated operators $E$ \cite{Ci00}. In
particular, one can use all the results obtained for the
separability of states \cite{separability}. This allows to
answer questions like "Given a CPM ${\cal E}$, can it be used to
create entanglement?''. Such questions may be relevant in
experiments, where one might want to know whether a certain
machine (set up) can be used to create entangled states.

\item %[{\bf (ii)}]
One can easily construct physical operations ${\cal E}$ which
can generate bound entangled states (BES).

\item %[{\bf (iii)}]
An important problem in the context of distributed quantum
computation \cite{Ci98} is the implementation of non--local
unitary operations. In \cite{Ci00}, it was shown that an
arbitrary 2--qubit unitary operation can be implemented using an
amount of entanglement which is proportional to the entanglement
capability of the operation \cite{Za00,Du00H}.
\end{itemize}

Then, we will discuss several other applications of the
isomorphism:

\begin{itemize}

\item %[{\bf (iv)}]
One can perform two--qubit gates {\it probabilistically} in the
context of single photon experiments via creation of entangled
states assisted by incomplete Bell measurements.

\item %[{\bf (v)}]
Several techniques concerning quantum states--- e.g. quantum
teleportation \cite{Be93T}, quantum state
purification\cite{Be93,Be96}, quantum data compression
\cite{Sc97} and quantum cloning \cite{cloning}--- were consider
in recent years. The isomorphism allows in a simple way to
obtain similar results for operations. That is, noisy unitary
operations can be purified, and sets of them can be stored and
compressed. Furthermore, it is possible to clone unitary
operations as well as to teleport them \cite{Hu00,Vi00}.
Finally, one can easily see how to perform the tomography
\cite{Po97,Ac00} of general non--local operations locally.

\item
One can perform certain non--local measurements by using a small
amount of entanglement.
\end{itemize}

This paper is organized as follows. In Sec. \ref{isomorph}, the
isomorphism between operations and states is reviewed and
several implications are discussed. This isomorphism provides
the basic tool for a number of applications which are presented
in the preceding sections. In Sec \ref{NLU}, we show how to
implement non--local two--qubit unitary operations with unit
probability, consuming an amount of entanglement which is
proportional to the entanglement capability of the operation.
Sec. \ref{purify} is concerned with purification of noisy
operations, while Sec. \ref{tom} deals with tomography of
arbitrary non--local operations. In Sec. \ref{Photo} it is shown
how to implement probabilistically two--qubit operations in the
context of single photon experiments. Next, in Sec. \ref{store},
storage and compression of non--local unitary operations are
discussed, while Sec. \ref{NLP} is concerned with the
implementation of non--local measurements. Finally, in Sec.
\ref{multi} the isomorphism is extended to multi--party systems.
We summarize our results in Sec. \ref{summary}.

%-------------------------------------------------------------------------
%-------------------------------------------------------------------------
\section{Isomorphism between operations and states}\label{isomorph}
%-------------------------------------------------------------------------
%-------------------------------------------------------------------------

In \cite{Ci00}, an isomorphism which relates physical operations
[equivalently completely positive maps (CPM) ${\cal E}$] acting
on two systems and (unnormalized) states (positive operators
$E$) was introduced. This isomorphism is an extension of the one
introduced by Jamiolkowski \cite{Ja75}. To be specific, let us
consider two spatially separated parties $A$ and $B$, each of
them possessing several particles\footnote{We will denote
particles belonging to party $A$ by $A_1,A_2,A_3$, where each of
the particles is a $d$--level systems. We will also use the
notation $A_{1,2}$ to refer to particles $A_1,A_2$. A similar
notation is used for particles belonging to party $B$.}. Let
$S=\{|i\rangle\}_{i=1}^d$ be an orthonormal basis, and
\bea
\label{MES} |\Phi\rangle_{A_{1,2}} &=&
\frac{1}{\sqrt{d}}\sum_{i=1}^d |i\rangle_{A_1}\otimes
|i\rangle_{A_2},\\
P_{A_{1,2}}&=&|\Phi\rangle_{A_{1,2}}\langle \Phi|,\label{P}
\eea
where $|\Phi\rangle$ is a maximally entangled state (MES) and
$P$ a projector on this state. We consider a CPM ${\cal E}$
acting on the density operator of two $d$--level systems, one
belonging to $A$ and one to $B$. Then, there exist an
isomorphism (linear one to one correspondence) between the CPM
${\cal E}$ and a positive operator $E$ \cite{Ci00} defined by
the following relations:
\bma\label{iso}\bea
E_{A_{1,2},B_{1,2}} &=& {\cal E}(P_{A_{1,2}}\otimes P_{B_{1,2}}),\label{iso1}\\
{\cal E}(\rho_{A_1B_1})&=& d^4{\rm tr}_{A_{2,3}B_{2,3}} %\nonumber\\&&
(E_{A_{1,2},B_{1,2}}
\rho_{A_3B_3} P_{A_{2,3}}P_{B_{2,3}})\label{iso2}.
\eea\ema

These equations have a very simple interpretation: On the one
hand, Eq. (\ref{iso1}) states that given a CPM ${\cal E}$, one
can always produce the state $E$ associated with ${\cal E}$ by
applying ${\cal E}$ to particles $A_1B_1$ if they are prepared
in the state $\tilde P=P_{A_{1,2}}\otimes P_{B_{1,2}}$. Note
that $\tilde P$ is a product state with respect to parties $A$
and $B$, while it is a local MES in the system belonging to
party $A$ and $B$ respectively. On the other hand, Eq.
(\ref{iso2}) says that given the state $E$ (of particles
$A_{1,2},B_{1,2}$), one can implement the operation ${\cal E}$
on an arbitrary state $\rho$ of two $d$--level systems
(particles $A_3B_3$) by measuring the projector $P$ locally in
$A_{2,3}$ and $B_{2,3}$. After a successful measurement ---where
the probability of success is given by $p=1/d^4$--- particles
$A_1B_1$ are found in the state ${\cal E}(\rho)$. In summary, a
CPM ${\cal E}$ can be used to prepare a state $E$, which in turn
can be used to implement ${\cal E}$ with a certain probability
of success.

%-------------------------------------------------------------------------
%-------------------------------------------------------------------------
\section{Implementation of non--local unitary operation with unit probability}\label{NLU}
%-------------------------------------------------------------------------
%-------------------------------------------------------------------------

In \cite{Ci00}, it was shown how to implement an arbitrary non--local two--qubit
unitary operation with arbitrary high accuracy and unit probability, consuming
an amount of entanglement which is proportional to the entangling capability of
the operation. Here, we review and improve this procedure. To this aim, ---as in
Ref. \cite{Ci00}--- we consider a family of phase gates
\be
U(\alpha_N) \equiv e^{-i\alpha_N \sigma_x^{A_1}\otimes\sigma_x^{B_1}},
\mbox{    }\alpha_N\equiv \pi/2^N\label{Ualpha}
\ee
where the $\sigma$'s are Pauli matrices. We show that:

\begin{description}
\item[{\bf (i)}] The operation $U(\alpha_N)$ can be implemented with probability $p=1/2$.
\item[{\bf (ii)}] By applying a finite sequence of operations of the form
(\ref{Ualpha}), each being implemented with probability $p=1/2$ using {\bf (i)},
one can achieve that the operation $U(\alpha_N)$ is applied with probability
$p=1$.
\item[{\bf (iii)}] Using gates of the form (\ref{Ualpha}) with binary angles
$\alpha_N=\pi/2^N$, one can implement phase gates with arbitrary angle $\alpha$.
\item[{\bf (iv)}] An arbitrary two--qubit unitary operation can be implemented using a
sequence of three operations of the form $U(\alpha)$, assisted by local unitary
transformations.
\end{description}
While {\bf (i-iii)} are already explained in \cite{Ci00}, the implementation of
{\bf (iv)} is different to the implementation described in \cite{Ci00}. There,
an infinite sequence of operations of the form $U(\alpha)$ was required in order
to implement an arbitrary two--qubit operation, while here a finite sequence
consisting of three operations suffices. The required amount of entanglement is
also smaller using the new method.

Since steps {\bf (i-iii)} will be crucial for the understanding of
some procedures described in later sections, we discuss them again
in detail. We start out by showing {\bf (i)}. First we note that the
operator associated with the unitary operation $U(\alpha_N)$
(\ref{Ualpha}) is given by $E_{A_{1,2},B_{1,2}}=
|\psi_{\alpha_N}\rangle_{A_{1,2},B_{1,2}}\langle \psi_{\alpha_N}|$, where
\bea
\label{psialpha} |\psi_{\alpha_N}\rangle_{A_{1,2},B_{1,2}} &=&\cos(\alpha_N)|\Phi^+\rangle_{A_{1,2}}|\Phi^+\rangle_{B_{1,2}}\nonumber\\
&& -i\sin(\alpha_N)|\Psi^+\rangle_{A_{1,2}} |\Psi^+\rangle_{B_{1,2}},
\eea
and $|\Phi^+\rangle=(|00\rangle+|11\rangle)/\sqrt{2},
|\Psi^+\rangle=\eins\otimes\sigma_x |\Phi^+\rangle =
(|01\rangle+|10\rangle)/\sqrt{2}$ are Bell states. In general, the
Bell basis is defined as
\be
\label{Bell} |\Psi_{i_1,i_2}\rangle=\eins\otimes \sigma_{i_1,i_2}|\Phi^+\rangle,
\ee
where $\sigma_{1,1}=\eins,\sigma_{1,2}=\sigma_x,\sigma_{2,1}=
\sigma_y$, $\sigma_{2,2}=\sigma_z$ and
$|\Psi_{1,1}\rangle=|\Phi^+\rangle,
|\Psi_{1,2}\rangle=|\Psi^+\rangle,
|\Psi_{2,1}\rangle=i|\Psi^-\rangle,
|\Psi_{2,2}\rangle=|\Phi^-\rangle$ are MES. Note that for
convenience---to ensure a simple notation below as well as in
the remaining sections---, we use a redundant definition of the
Bell basis. We consider a situation similar to the one described
in (\ref{iso2}); that is, particles $A_{1,2}B_{1,2}$ are
prepared in state (\ref{psialpha}), and $\rho_{A_3B_3}$ is the
state on which a CPM ${\cal E}$ --- in our case, the unitary
operation $U(\alpha_N)$ (\ref{Ualpha}) --- should be applied.
Now, a Bell measurement is performed on particles $A_{2,3}$
[$B_{2,3}$]. Assume that the result associated to
$|\Psi_{i_1,i_2}\rangle$ [$|\Psi_{j_1,j_2}\rangle$] is obtained.
In this case, the state of system $A_1B_1$ is proportional to
\be
{\cal E} \left((\sigma_{i_1,i_2}^{A_1}\otimes\sigma_{j_1,j_2}^{B_1}) \rho_{A_1B_1}
(\sigma_{i_1,i_2}^{A_1}\otimes\sigma_{j_1,j_2}^{B_1})\right).
\ee
Thus, as a result of the measurement we either implement the CPM
${\cal E}$, or some unitary operation followed by the CPM. We
now proceed as follows: In case the result of the measurement
was $i_1,i_2$ [$j_1,j_2$], the local unitary operation
$\sigma_{i_1,i_2}$ [$\sigma_{j_1,j_2}$] is applied on $A_1$
[$B_1$]. In case that ${\cal E}$ is given by the unitary
operation $U(\alpha_N)$ (\ref{Ualpha}), one readily observes
that the resulting operation performed on $\rho_{A_1B_2}$ after
this procedure will be (i) $U(\alpha_N)$ if $i_1=j_1$; (ii)
$U(\alpha_N)^\dagger=U(-\alpha_N)$ if $i_1\ne j_1$. Due to the
fact that all measurement outcomes are equal probable, we have
that with probability $p=1/2$ the desired operation
$U(\alpha_N)$ was applied, while with $p=1/2$ the operation
$U(-\alpha_N)$ was performed, from which {\bf (i)} follows.

Before we proceed, we investigate the amount of non---local entanglement
(between systems $A$ and $B$)  which is required to perform the described
procedure. The amount of entanglement of the state $|\psi_{\alpha_N}\rangle$
(\ref{psialpha}) is given by its entropy of entanglement
\be
E(\psi_{\alpha_N})=-x_N
\log_2(x_N)-(1-x_N)\log_2(1-x_N),\label{EOE}
\ee
where $x_N=\cos^2(\alpha_N)=\cos^2(\pi/2^N)$. That is, the
amount of entanglement required to implement the operation
$U(\alpha_N)$ with probability $p=1/2$ is given by (\ref{EOE}).
We have that $U(\pi/2)=-i \sigma_x\otimes \sigma_x$ is a {\it
local} gate and thus $E(\psi_{\alpha_1})=0$, while
$E(\psi_{\alpha_2})=1$, i.e. one ebit of entanglement is
required. For $N\geq 2$, we have that $E(\psi_{\alpha_N})$ is
monotonically decreasing with $N$. The amount of classical
communication is given by one bit in both directions (the value
of $i_1$ [$j_1$] respectively has to be transmitted).

Regarding {\bf (ii)}, we have to show how to obtain a
probability of success $p=1$ by making use of the procedure
described above. Note that with probability $p=1/2$, we succeed
and apply the desired gate, while with $p=1/2$ we fail and apply
$U(-\alpha_N)$ instead. Now, if we fail, we repeat the procedure
but with systems $A_{1,2}B_{1,2}$ prepared in the state
$|\psi_{2\alpha_N}\rangle$. With a probability 1/2 we succeed,
and otherwise we will have applied $U(-\alpha_N)^3$ to the
original state instead. We continue in the same vain, that is in
the $k$--th step we use systems $A_{1,2}B_{1,2}$ prepared in the
state $|\psi_{2^{k-1}\alpha_N}\rangle$ so that if we fail
altogether we will have applied $U(-\alpha_N)^{2^{k}-1}$. For
$k=N$ we have that $U(-\alpha_N)^{2^{N}-1}=-U(\alpha_N)$, and
therefore even if we fail we will have applied the right gate,
so that the procedure ends. In fact, the $N^{\rm th}$ step will
succeed with $p=1$, as $U(\pi/2)$ is a local gate which can be
implemented with unit probability and without consuming
entanglement. That is, a sequence of $N$ operations of the form
(\ref{Ualpha}) allows to implement the operation $U(\alpha_N)$
with unit probability, which proves {\bf (ii)}.

Let us investigate the average amount of entanglement which is consumed during this procedure.
We have
\be
\label{avent}
\bar E[U(\alpha_N)] = \sum_{k=1}^{N} \left(\frac{1}{2}\right)^{k-1}E(\psi_{\alpha_{N-k+1}})
= \alpha_N f_N,
\ee
where
\be
f_N = \frac{1}{\pi} \sum_{k=1}^{N} 2^k E(\psi_{\alpha_{k}})< f_{\infty}=5.97932.
\ee
In (\ref{avent}), the weight factor $p_k=(1/2)^{k-1}$ gives the probability
that the $k$--th step has to be performed. Thus, we obtain $\bar E[U(\alpha_N)]<
\alpha_N f_\infty$; that is, the average amount of entanglement is bounded from
above by a quantity which is proportional to the angle $\alpha_N$ and thus
---for small $\alpha_N$--- proportional to the entangling capability of the
operation \cite{Du00H}. The average amount of classical communication is given
by $2-(1/2)^{N-2}$ bits.

To show {\bf (iii)}, we use the fact that any gate $U(\alpha)$
with arbitrary phase $\alpha$ can be approximated with arbitrary
high accuracy by a sequence of gates of the form $U(\alpha_N)$.
That is, any angle $0\leq\alpha\leq \pi$ can be written as
\be
\alpha=\pi\sum_{k=1}^{\infty} n_k 2^{-k}, \mbox{   } n_k \in \{0,1\}\label{binalpha}.
\ee
For each $k$, we have that $n_k$ is either ``0'' ---which means that the
rotation $U(\alpha_k)$ does not have to be performed---  or ``1'' ---which means
that the rotation $U(\alpha_k)$ has to be performed---. Operations of the form
$U(\alpha_k)$ can be implemented with unit probability using {\bf(i-ii)}. The
average amount of entanglement consumed to implement $U(\alpha)$ is bounded by
$\bar E \leq f_\infty \alpha$ ebits.

Finally, to show {\bf(iv)}, we use the result of Kraus {\it et al.} \cite{Kr00}.
There, it was shown that an arbitrary two--qubit unitary operation can be written in the
following form
\be
U_{AB}=V\otimes W e^{-i H} \tilde V \otimes\tilde W,
\ee
where $V,W,\tilde V,\tilde W$ are local operations and
\be
H=\sum_{k=x,y,z}^3 \mu_k \sigma^A_k \otimes \sigma^B_k \equiv \sum_{k=1}^3 H_k,
\ee
where $0\leq \mu_k\leq\pi/2$. We note that
\be
e^{-i H}=e^{-i H_1}e^{-i H_2}e^{-i H_3},
\ee
and $e^{-i H_k}$ are ---up to a change of local basis---
operations of the form (\ref{Ualpha}) for which we already
provided a protocol [see {\bf (i-iii)}]. Using this, we obtain
that an arbitrary two--qubit unitary operation can be performed
using a sequence of three operations of the form $U(\alpha)$,
assisted by local unitary operations, which proves {\bf (iv)}.
The required amount of entanglement is bounded by $f_\infty
(\mu_1+\mu_2+\mu_3)$ ebits.

%-------------------------------------------------------------------------
%-------------------------------------------------------------------------
\section{Purification of noisy operations}\label{purify}
%-------------------------------------------------------------------------
%-------------------------------------------------------------------------

 %State problem:

In this Section, we consider purification of a noisy operations. We will discuss
two different scenarios:

In the first scenario, we consider two spatially separated
parties $A$ and $B$ who want to perform a {\it known,
non--local} (entangling) unitary operation $U$ between two
particles they share. We will assume that $A$ and $B$ are only
able to perform the operation $U$ in an imperfect way. So
instead of performing $U$ on their particles, they perform some
CPM ${\cal E}_U$. The problem we pose is the following: Given
several applications of the noisy operation ${\cal E}_U$ and
arbitrary local resources, can the parties $A$ and $B$ use them
to perform the (noiseless) operation $U$ on an arbitrary state
of two qubits instead? Under which circumstances is this
possible? In case this is possible, we say that the noisy
operation is purificable. In Sec. \ref{known} we are going to
show when and how it is possible to achieve this task.

The second scenario is concerned with the purification of an {\it unknown,
local} noisy unitary operation ${\cal E}_U$,  where we explicitly assume a
specific form of noise.  In Sec. \ref{unknown}, we provide a procedure to
implement an unknown unitary operation perfectly, given several applications of
the noisy operation.

In both cases, it turns out that the isomorphism (\ref{iso})
allows to use results obtained for purification of states and
thus for a very simple solution to the problem. Regarding the
first scenario, the corresponding problem for states is the
problem of entanglement distillation of mixed states
\cite{Be96}. For the second scenario, the corresponding problem
for states is the purification of a single qubit \cite{Ci99}.

\subsection{Purification of a known non--local noisy unitary operation}\label{known}

We consider two parties $A$ and $B$, who want to perform a joint unitary
operation $U$ among two particles they share. For simplicity, let us assume that
$U\in SU(4)$, i.e. the particles are qubits. The parties $A$ and $B$ are only
capable to perform the operation $U$ in an imperfect way, so they perform
some CPM ${\cal E}_U$ instead. For example, a noisy $N$--qubit operation can be of
the form \cite{note1}
\be
{\cal E}_U(\rho)=qU\rho
U^\dagger+\frac{1-q}{2^N}\eins,\label{noisyU}
\ee
i.e. with probability $q$ the desired operation is performed, while with $(1-q)$
a completely depolarized state (described by the identity operator $\eins$) is
produced. The following analysis is not restricted to this specific
form of noisy operations.

The operation $U$ is known to both $A$ and $B$. Furthermore,
they are allowed to use auxiliary systems and are able to
perform all operations (including two--qubit operations) on
their individual sites perfectly. In the following, we are going
to show that the noisy, entangling operation ${\cal E}_U$ can be
purified if and only if the operator $\rho_{\cal E}$
corresponding to ${\cal E}_U$ (\ref{iso1}) is distillable. We
also provide a practical protocol to achieve this task.

The purification procedure takes place as follows:
\begin{itemize}
\item ${\cal E}_U$ is used to create several copies of $\rho_{\cal E}$ (see (\ref{iso1})).
\item With help of entanglement distillation for states, out of $\rho_{\cal
E}^{\otimes M}$ a number of MES are created.
\item The MES are used to create a set of states of the form (\ref{psialpha}),
either via deterministic state transformation (single copy case) or via
entanglement dilution \cite{Be93}.
\item Finally, these states are used to implement $U$ with unit probability and
arbitrary high accuracy as described in Sec. \ref{NLU}.
\end{itemize}

Now we will show that an operation ${\cal E}_U$ ---where $U$ is
an entangling operation--- is purificable if and only if
$\rho_{\cal E}$ is distillable. This can be seen as follows. On
the one hand, if $\rho_{\cal E}$ is distillable, one can use the
procedure described above to purify the noisy operation ${\cal
E}_U$. On the other hand, if ${\cal E}_U$ is purificable, this
implies that the unitary operation $U$ can be performed on an
arbitrary state of two qubits, using a sequence of noisy
operations ${\cal E}_U$ assisted by local operations and
classical communication. Since $U$ is an entangling operation,
the corresponding pure state $E_U$ is also entangled. That is,
the sequence of operations ${\cal E}_U$, assisted by local
operations and classical communication is capable to create
entangled states when acting on a certain separable state. Using
the isomorphism (\ref{iso}), we can write this sequence of
operations acting on a separable state in terms of a trace over
several operators $E_i$, where local operations in the sequence
correspond to separable operators \cite{Ci00}. That is, the only
entangled operators which appear in this expression are
operators $\rho_{\cal E}$ corresponding to the noisy operation
${\cal E}_U$ and the resulting state is entangled. This implies
that from several copies of the mixed state $\rho_{\cal E}$, an
entangled pure state can be created. Note that using
entanglement distillation for pure states \cite{Be93}, this
implies that one can also create a MES. We thus have that
$\rho_{\cal E}$ is distillable, which finishes the proof of our
statement.

Since ${\cal E}_U$ is a general CPM, $\rho_{\cal E}$ is a mixed
state in $\C^4\otimes\C^4$, where no necessary and sufficient
condition for distillability is known (see however
\cite{Du00D,Di00}). It is known that non--positive partial
transposition of $\rho_{\cal E}$ is a necessary condition for
distillability, however there are strong evidences that this
condition is not sufficient \cite{Du00D,Di00}. Using
entanglement purification for states, e.g. via the methods
discussed in \cite{Du00D,Di00,Ho98D,Be96}, one may be able to
obtain a MES starting from several copies of $\rho_{\cal E}$.

Given the error model (\ref{noisyU}), one can obtain a necessary
and sufficient condition for purificability. It turns out that
unitary operations which are only weakly entangling (e.g.
operations of the form $U(\alpha)$ with $\alpha \ll 1$) are much
more sensitive to noise than unitary operations which are
strongly entangling, e.g. the controlled--not (CNOT)
operation\cite{noteCNOT}. This means that the tolerable error,
specified by $(1-q)$ ---such that purification of the noisy
operation is still possible--- in the case of the CNOT is much
bigger than for $U(\alpha)$ with $\alpha \ll 1$. For the CNOT
and the error model (\ref{noisyU}), one obtains $q > 1/9$ in
order that purification of the noisy gate be possible
\cite{note2}. For operations of the form $U(\alpha)$, one finds
$q>(16\cos(\alpha)\sin(\alpha)+1)^{-1}$ as a necessary and
sufficient condition that gate purification is possible. For
$\alpha=\pi/2^{13}$, this value is e.g. given by $q > 163/164
\approx 0.994$, i.e. less than one percent of noise is allowed
in this case.

Note that the process of entanglement distillation involves
two--qubit joint operations as well. The reason why we treat
these (local) operations differently than the (non--local)
operation $U$ can be viewed as follows. On the one hand, the
parties $A$ and $B$ may be spatially separated and the
interaction between the two parties ---for example performed
through the usage of a (noisy) quantum channel--- is much more
sensitive to noise than the local operations performed by only
one of the parties. On the other hand, each of the parties $A$
and $B$ may be considered to possess a single particle only,
each particle containing several levels. Here, the additional
levels are used instead of the auxiliary qubits. In this case,
the operation $U$ is concerned with the interaction between two
different particles, while all local operations (also including
multilevel --- equivalently multiqubit--- operations) are
operations performed on a single particle, which are much easier
to implement. For example, using atoms or ions with several
levels, all local operations can be easily performed
\cite{ionsatoms}. However, controlled interactions between two
ions/atoms are very difficult to achieve, which leads to the
fact that two--particle gates are noisy while local gates are
practically not. Recall that in state purification, it is
similarly assumed that local operations can be performed
perfectly and that it has to be known which is the MES which has
a large overlap with the mixed state the parties share in order
that they can distill this specific MES. Similarly, knowledge of
the perfect unitary operation $U$ is required.

One may also consider that the local operations are noisy. In
this case, both the process of distillation of states and the
implementation of the operation $U$ using several different
states and Bell measurements will give rise to some
imperfections. The purification of states with imperfect means
was studied in \cite{Br98,Gi98} and it was found that no MES can
be obtained if the local operations are noisy and a certain
error model is assumed. However, one is still able to increase
the fidelity, i.e. the overlap of the produced state with a MES,
where the maximal reachable fidelity is determined by the amount
of noise introduced by the local operations. So, instead of
producing MES, one produces some mixed state $\rho$. This state
$\rho$ may then be transformed to a state which is close to the
ones of the form (\ref{psialpha}). These states can then be used
to implement the operation $U$ in an imperfect way, since both
the states which are used and the operations which are performed
are noisy. Furthermore, one should take into account that a
sequence of noisy operations is required in order to implement
$U$ with unit probability, so the errors may accumulate. For
almost perfect local operations and very noisy non--local
operations, one may, however, still expect a purification
effect.

\subsection{Purification of an unknown local noisy unitary operation}\label{unknown}

Here, we consider a party $A$ who wants to perform a unitary
operation $U \in SU(2)$ on a single qubit. The operation cannot
be performed perfectly but is subjected to some noise. We will
explicitly assume that the imperfect operation ${\cal E}_U$ is
of the form (\ref{noisyU}) \cite{note1} with $N=1$; that is,
with probability $q$ the desired unitary operation $U$ is
performed, while with probability $(1-q)$ the completely
depolarized state $1/2\eins$ is produced. Here, in contrast to
the previous discussion, we will assume that the operation is
local and {\it unknown} to $A$ (for example, ${\cal E}_U$ is
provided to $A$ by a second party via a black box). Given that
party $A$ is able to perform all operations perfectly, we will
show that the unknown noisy unitary operation can be purified,
i.e. via several applications of ${\cal E}_U$, the noiseless
operation $U$ can be implemented on an arbitrary qubit. For
simplicity, we assume that the unitary operation $U$ is of the
form
\be
U(\alpha)=e^{-i\alpha\sigma_x},\label{Ua}
\ee
where $\alpha$ is unknown, however the analysis can be generalized to arbitrary
single--qubit unitary operations. The positive operator $E$ corresponding to the imperfect
operation ${\cal E}_U$ is given by
\be
E=q|\Psi_U\rangle\langle\Psi_U|+\frac{1-q}{4}\eins_4,
\ee
where $|\Psi_U\rangle=\cos(\alpha)|\Phi^+\rangle - i \sin(\alpha)
|\Psi^+\rangle$.

We proceed as follows. First, we project $E$ on the subspace spanned by
$\{|\Phi^+\rangle,|\Psi^+\rangle\}$ and relabel the basis:
\be
|\tilde 0\rangle=|\Phi^+\rangle, \mbox{   } |\tilde 1\rangle=-i|\Psi^+\rangle.\label{bc}
\ee
In case we succeed,
which happens with probability $(q+1)/2$, the resulting state will be
\be
\tilde E =\lambda |\tilde \Psi_U\rangle\langle \tilde \Psi_U|+ (1-\lambda) \frac{1}{2}\eins_2, \label{EU}
\ee
where $|\tilde \Psi_U\rangle=\cos(\alpha)|\tilde 0\rangle + \sin(\alpha)|\tilde
1\rangle$ and $\lambda\equiv(2q)/(1+q)$. Given $N$ states of the form
(\ref{EU}), one can use the procedure described in \cite{Ci99} to purify the
noisy state, i.e. to increase $\lambda$. For large $N$, the average fidelity
---that is the overlap of the produced states with the state
$|\tilde\Psi_U\rangle$--- scales like $F \approx
1-\frac{1}{2N}\frac{1-\lambda}{\lambda^2}$, whereas the yield
---i.e. fraction of the number of produced states to the number of
initial states $N$--- scales like
$D\approx\lambda+\frac{1}{N}\frac{1-\lambda}{\lambda}$
\cite{Ci99}. That is, for $N\rightarrow \infty$ one obtains almost
perfect states $|\tilde\Psi_U\rangle$ with a yield $\lambda$. Note
that the states $|\tilde\Psi_U\rangle$ are not uniformly
distributed on the whole Bloch's sphere but rather only on the
equatorial plane. Nevertheless, one can still use the same
procedure as described in \cite{Ci99}, where a uniform
distribution was assumed. The corresponding values for $F$ and $D$
in our case are at least as big as the ones obtained in
\cite{Ci99}, since we have additional knowledge of the state,
which may be used to further increase $F$ and $D$.
Note that in order that purification is possible, we need that $\lambda>0$ and
thus $q>0$. So all noisy gates of the form (\ref{noisyU}) and $U$
given by (\ref{Ua}) can be purified if $q>0$.

To summarize, we managed to produce an arbitrary number of (almost) perfect
states $|\tilde\Psi_U\rangle$ given several applications of the noisy operation
${\cal E}_U$. Note that $|\tilde\Psi_U\rangle$ can be transformed
deterministically to $|\Psi_U\rangle$ by undoing the basis change (\ref{bc}).
From the results of Sec. \ref{NLU}, we know that the state $|\Psi_U\rangle$ can
be used to implement $U$ with probability $p=1/2$. What remains is to show that
one can implement $U$ with probability $p=1$. The simplest way to see this is
the following: If we fail, we try to implement $U$ again, we make a third
attempt and try to implement $U$ and so on. Every odd number of steps, say
$2j+1$, we stop the procedure if we have succeeded in $(j+1)$ steps and did not
succeed in $j$ steps. In this case, we have applied the operation $U$ in total
$(j+1)$ times and the operation $U^\dagger$ $j$ times, which is equivalent to
apply the operation $U$. This is a one--side bounded random walk with
probability $p=1/2$, where one can easily see that the total success probability
converges to $p=1$. Alternatively, one can also use the operation $U$ to prepare
states $|\Psi_U\rangle$ with coefficient $2^k\alpha$, which is possible with
probability $p=1/2^{2^k}$. These states can than be used to implement $U$ with
$p=1$ following the procedure described in \cite{Vi00}. For a success
probability $p=1-o(\epsilon)$, in total $o(\epsilon^{-1})$ states
$|\Psi_U\rangle$ with coefficient $\alpha$ are required.

Alternatively to the procedure described above, one may also use a method
similar to the one of Sec. \ref{tom} to implement $U$ given several applications
of ${\cal E_U}$. By a sequence of measurements one first determines the
state $E$, from which $|\Psi_U\rangle$ can be found and used to implement the
operation $U$ (which is now known to $A$).

%-------------------------------------------------------------------------
%-------------------------------------------------------------------------
\section{Tomography of operations}\label{tom}
%-------------------------------------------------------------------------
%-------------------------------------------------------------------------

In this Section, we consider the problem of tomography of an arbitrary, unknown
non--local CPM. Given several applications of the unknown CPM ${\cal E}$ and
using the isomorphism (\ref{iso}), it is straightforward to completely determine
the non--local CPM by a sequence of {\it local} measurements assisted by
classical communication. To this aim, we use the operation ${\cal E}$ to prepare
several copies of the associated state $E$ (\ref{iso1}). Now, using tomography
for states \cite{Po97}, the state $E$ ---and thus via (\ref{iso2}) also the CPM $\cal{E}$---
can be determined.

Next we show that a sequence of local measurements assisted by classical
communication suffices to completely determine a non--local mixed state $E$ (and
thus a non--local CPM ${\cal E}$). Let $A$ and $B$ be two spatially separated
parties and $\{A_i\}$ [$\{B_j\}$] be an orthonormal \cite{ortho} basis of
self adjoint operators in $A$ [$B$]. We have that $E_{AB}$ can be
written as
\be
E_{AB}=\sum_{i,j} \lambda_{ij} A_i\otimes B_j,
\ee
where $\lambda_{ij}={\rm tr}(A_i\otimes B_j E_{AB})$ is the expectation value of
the operator $A_i\otimes B_j$. Now, by measuring the operators $A_i$ [$B_j$]
locally in $A$ [$B$] and using classical communication, one can
establish the values of all $\lambda_{ij}$ and thus the state $E_{AB}$. In case
the operation ${\cal E}$ acts on two qubits, the corresponding state $E_{AB}$ is
a state of two four--level systems. The set of operators $\{A_i\}$ can e.g.
chosen to be $\{\sigma_{i_1,i_2}\otimes\sigma_{i_3,i_4}\}$ ---where $\sigma_{i_1,i2}$ are
defined in Sec. \ref{NLU} (see Eq. (\ref{Bell}))--- and similarly for $\{B_i\}$.

%-------------------------------------------------------------------------
%-------------------------------------------------------------------------
\section{Probabilistic implementation}\label{Photo}
%-------------------------------------------------------------------------
%-------------------------------------------------------------------------

In this Section, we will show that the possibility to distinguish the state
$|\Phi^+\rangle$ from the other three Bell states and the capability to produce
certain entangled states allows to implement {\it probabilistically} arbitrary two
particle unitary operations. This has applications in the context of single
photon experiments, since our method allows to implement two photon gates with a
certain probability of success, which is already sufficient to implement
entanglement distillation. Note that this should be feasible even with present
day technology.

In the following, we concentrate on two qubit gates. Given the
results of Sec. \ref{NLU}, one observes that the possibility of
creating certain entangled states, together with the capability of
performing local Bell measurements allows to implement an
arbitrary 2--qubit operation \cite{Go99}. That is, the problem to perform
two--qubit gates is shifted to the problem of
\begin{description}
\item[{\bf (i)}] creating certain entangled states and
\item[{\bf (ii)}] the capability to perform perfect Bell measurements.
\end{description}
In the following, we will discuss {\bf (i-ii)} in the context of single photon
experiments.

Regarding {\bf (i)}, in single photon experiments one is already
able to create certain MES (e.g. via parametric down conversion).
For example, MES of two qubits were created and used in
teleportation experiments \cite{photons}. In addition, the
creation of a three qubit Greenberger-Horne-Zeilinger (GHZ) state
was reported \cite{GHZ}. Although non--linear elements are
required in order to produce entangled states, it is much easier
to use these elements in such a way that a {\it known} state is
generated rather than using some non--linear elements to perform a
controlled interaction between arbitrary states.
Applying the isomorphism (\ref{iso}), one observes that the state
$E$ corresponding to a general two--qubit unitary operation is a
pure state of two four level systems (equivalently of four
qubits). For example, the state corresponding to the CNOT
operation \cite{noteCNOT} is given by $E_{\rm
CNOT}=(|00\rangle_A|\Phi^+\rangle_B+|11\rangle_A|\Psi^+\rangle_B)/\sqrt{2}$,
while the SWAP operation (which is given by the mapping
$|ij\rangle\rightarrow |ji\rangle$) is specified by $E_{\rm
SWAP}=|\Phi^+\rangle_{A_1B_2}|\Phi^+\rangle_{A_2B_1}$. Operations
of the form (\ref{Ualpha}) are specified by states of the form
(\ref{psialpha}). Due to the fact that the states
(\ref{psialpha}), as well as $E_{\rm CNOT}$ only have two Schmidt
coefficients (when considered as a bipartite system $A-B$), it
should be possible to create them in the laboratory using present
day technology.

What remains is {\bf (ii)}, the problem to perform Bell
measurements. For single photons, using non--linear elements
only (beam splitters and photo detectors), one is able to
perform {\it incomplete} Bell measurements. In particular, one
can perfectly distinguish the three sets of states
$\{|\Phi^+\rangle\}$, $\{|\Psi^+\rangle\}$, and
$\{|\Phi^-\rangle,|\Psi^-\rangle\}$ \cite{BellM}. The optimality
of this process using linear elements was discussed in
\cite{Lu99}. Due to the fact that Bell measurements cannot be
performed perfectly with linear elements \cite{Lu99} (see
however \cite{Kn00}), it follows that two--qubit gates cannot be
implemented with unit probability using the procedure described
in Sec. \ref{NLU}. However, even incomplete Bell measurements
(which can already be performed in the laboratory) still allow
for a {\it probabilistic} implementation of arbitrary two qubit
gates. That is, with a certain probability the desired gate is
applied, while otherwise a different (possibly unknown)
operation is performed. In the latter case, the input state has
to be discarded.

Let us investigate the consequences of incomplete Bell
measurements a bit closer. From (\ref{iso}), we know that if
both parties $A$ and $B$ obtain as a measurement outcome the
state $|\Phi^+\rangle$, the desired operation was performed. Due
to the fact that $|\Phi^+\rangle$ can be perfectly distinguished
from the other three Bell states using the methods described in
\cite{BellM,Lu99}, and all measurement outcomes are equal
probable (in the case of two qubits, $p_{\Phi^+}=1/4$), this
allows to implement the desired unitary operation with
probability $p=1/16$. For unitary operations of the form
(\ref{Ualpha}), this probability can be further increased to
$p=1/4$ given the fact that also $|\Psi^+\rangle$ can be
perfectly distinguished from the other Bell states. That is, if
both party $A$ and $B$ find as a measurement outcome either
$|\Phi^+\rangle$ or $|\Psi^+\rangle$, the desired unitary
operation was performed. In case the outcome was
$|\Psi^+\rangle$, additional application of the local operation
$\sigma_x$ is required (see Sec. \ref{NLU}).

Note that probabilistic implementation of two--qubit gates is
not useful in the context of quantum computation, as
probabilistic operations may change the complexity class of the
problem and may thus destroy the (exponential) speed up of the
quantum algorithm in question. However, probabilistic gates are
useful for processes such as entanglement distillation
\cite{Be96}, which itself is already a probabilistic process.
For example, this may help in the implementation of quantum
repeaters \cite{Br98} using photons only (i.e., for quantum
communication over arbitrary distances). Due to the fact that
photons are ideal candidates for quantum communication (due to
their fast propagation), it is highly desirable to manipulate
them directly (e.g. to perform entanglement purification as
required in the quantum repeater protocol \cite{Br98}) rather
than mapping their states on the states of another physical
system, e.g. of an ion or an atom, and vice versa. The method
discussed in this section may help to achieve this task.

Recently, an alternative approach was presented by Pan et. al
\cite{Pa00}, where entanglement purification without CNOT
operations was discussed. As this approach is concerned with a
certain distillation procedure only, the solution provided in
Ref. \cite{Pa00} to this specific problem is more efficient than
the one we obtain here. However, we provide a more general
framework which allows to implement {\it arbitrary} two--quibit
operations probabilistically. Another proposal was presented by
Knill et. al \cite{Kn00}, who showed the implementation of a
certain two--qubit operation with unit probability, taking full
usage of all resources (i.e. using an arbitrary number of
modes).

Note that similar techniques may be used to speed up ---in some
sense--- slow two--particle interactions. The scenario we have
in mind is the following: At a certain time ---e.g. in course of
a quantum computation--- an entangling quantum operation should
be performed on two particles. If the interaction between two
particles is weak, the required interaction time in order that a
entangling operation can be performed will be large. Now,
instead of performing the operation when it is required, we use
the (slow) interaction at some earlier stage to prepare certain
entangled states. These states can then ---at later time--- be
used to implement the two--particle operation almost immediately
---once the two particles on which the operation should be
performed arrive--- using the procedure described in Sec.
\ref{NLU}. Although this procedure involves local Bell
measurements, this will not slow down the process as for the
implementation of those measurements, no two--particle
interactions are required. For example, we can use internal
levels of atoms or ions instead of local auxiliary qubits (see
also Sec. \ref{purify}). Bell measurements in this case involve
only single particle interactions between the different levels
of a particle, which we assumed to be much faster than
two--particle interactions.

%-------------------------------------------------------------------------
%-------------------------------------------------------------------------
\section{Storage and compression of unitary operations}\label{store}
%-------------------------------------------------------------------------
%-------------------------------------------------------------------------

 % State 3 problems:    1) local storage of unitary operations
 %          2) non--local unitaries: non--local storage
 %          3) non--local unitary operations: local storage + non--local implementation

In this Section, we will discuss storage \cite{Ni97,Vi00} and
compression of unitary operations. We consider a (possibly
infinite) set of unitary operations $U_1,U_2,\ldots,U_N$. Each
operation is assigned an a priori probability $p_i$. We consider
a long sequence of those operations, where each element $U_i$ of
this sequence is chosen at random according to the probability
distribution $\{p_i\}$. We are interested in the average number
of qubits which are required to store one of the operations
$U_i$ and implement the operations at later time with unit
probability and high accuracy. We consider the following
variations of this problem:

\begin{description}
\item[{\bf (i)}] The operations $U_i$ are {\it local}.
That is, a party $A$ locally stores a certain number of qubits
and uses these qubits to implement one of the local operation
$U_i$ on some unknown state at later time. In this case, we are
interested in the average number of qubits to be stored locally.

\item[{\bf (ii)}] The operations $U_i$ are {\it
non--local}. That is, two spatially separated parties $A$ and
$B$ store a set of (possibly entangled) states and use these
states later on to implement the non--local operation $U_i$. In
this case, we allow the parties $A$ and $B$ to share some
initial entangled states. The storage procedure, however, is
restricted to local operations only. That is, parties $A$ and
$B$ store (and compress) their part of the system individually.
We are interested in the average number of qubits required in
$A$ ($B$) to store one of the operations $U_i$ locally.

\item[{\bf (iii)}] The operations $U_i$ are {\it non--local}. In
contrast to {\bf (ii)}, one of the parties, say $A$, stores the operations {\it
locally}. Using quantum communication, part of the stored system is then
transferred to $B$ and finally used to implement the non--local operation
$U_i$. In this case, we are interested in the average number of qubits which have
to be stored locally in $A$ as well as in the required quantum communication,
i.e. the average number of qubits which have to be transmitted from $A$ to $B$.
\end{description}

Note that in all cases, the unitary operation to be performed is at any stage
unknown to $A$ (and $B$). We will show that storage of certain sets of unitary
operations is possible. Furthermore, the scheme we propose allows to compress
the amount of required storage qubits (as well as the amount of qubits
transmitted from $A$ to $B$ in {\bf (iii)}) if one restricts the set of allowed
operations to a certain subset. It turns out that even for an infinite set of
operations $U_i$, the average amount of required storage qubits per operation
can be much smaller than one. These results can be viewed as an extension of the
Schumacher data compression for states \cite{Sc97} to unitary operations. In
fact, we will use the results of Ref. \cite{Sc97} to achieve this task.

Very recently, the problem of storage of a general unitary
operation was considered by Vidal {\it et al.} \cite{Vi00} and
an optimal solution was provided. In contrary to Ref.
\cite{Vi00}, we propose schemes which are capable to compress
the required amount of storage qubits and also discuss storage
of non--local operations. We will propose two different schemes
for storage, one dealing with a possible infinite set of unitary
operations $U_i$ and one with a finite set. We will discuss both
schemes in the context of {\bf (i-iii)}.

%-------------------------------------------------------------------------
%-------------------------------------------------------------------------
\subsection{Local storage of local unitary operations}\label{loc}

We start out with {\bf (i)}, the local storage of a set of local
unitary operations. We consider unitary operations acting on two
qubits and assume that they are local, i.e., both qubits on
which the operation should be performed are held by the same
party, say $A$.

\subsubsection{Storage of an infinite set of unitary operations}\label{inf}

Here, we describe a procedure to store {\it locally} a unitary operation of the form
$U(\alpha)$ (\ref{Ualpha}) with arbitrary, unknown $\alpha$ using on average
less than four [0.2257] qubits per operation if $0\leq\alpha\leq\pi$
[$\pi/8$].
We assume uniform distribution of angles $\alpha$, i.e. any operation is
equally likely.

We remind the reader that an operation $U(\alpha)$ for arbitrary
$\alpha$ can be implemented by a sequence of operations of the
form $U(\alpha_k)$ (\ref{Ualpha}) with binary angles
$\alpha_k=\pi/2^k$ (see Sec. \ref{NLU} {\bf (i-iii)}). Using the
fact that $\alpha$ can be written in binary notation
(\ref{binalpha}) and assuming that all angles are equally
likely, it follows that $n_k=0$ and $n_k=1$ are equal likely
$\forall k$.

We first consider the implementation of $U(\alpha_k$) for a
certain $k\equiv N$ and $\alpha_N=\pi/2^N$. Following the procedure described in
Sec. \ref{NLU} {\bf (ii)}, we have that if $n_N=1$, the following set of $N$ states
is required to implement this operation with probability $p=1$:
\be
G_N=\{|\Psi_{\alpha_N}\rangle,|\Psi_{2\alpha_N}\rangle,\ldots,|\Psi_{2^{N-1}\alpha_N}\rangle\},\label{S_N}
\ee
where the corresponding probabilities are given by $p_l=1/2^{l-1}$ for the
$l^{\rm th}$ state. If however $n_N=0$, no operation has to be performed. In
this case, one can store the set of states $\tilde
G_N=\{|\Psi_{0}\rangle,|\Psi_{0}\rangle,\ldots,|\Psi_{0}\rangle\}$, which
corresponds to the implementation of the identity operation in each step.
However, each step can be considered independently and involves with probability $p=1/2$ either the
storage of the state $|\Psi_{2^l\alpha_N}\rangle$ for the $l^{\rm th}$ step if $n_N=1$ or
$|\Psi_{0}\rangle$ if $n_N=0$. Thus, one can us data compression of pure states \cite{Sc97} for
each step independently. The corresponding compression factor $S_j$ for the $l^{\rm th}$ step is given by
the entropy of the operator $\tilde\rho$, which is an equal mixture of the state
$|\Psi_{\alpha_j}\rangle$ and $|\Psi_{0}\rangle$, where $j\equiv (N-l)$. One finds
\be
S_j=-x_j \log_2(x_j)-(1-x_j)\log_2(1-x_j),
\ee
with $x_j=(1+\cos{\alpha_j})/2$ and $\alpha_j=\pi/2^j$. Recall also that the
$l^{\rm th}$ step has to be performed only with probability $p_l$. That is, the
total amount of qubits required to store the operation $U(\alpha_N)$, where it is
unknown whether it should be performed or not, is given by
\be
\sum_{l=1}^{N} S_{N-l}\frac{1}{2^{l-1}} \label{Sl}.
\ee

We now consider a sequence of operations of the form $U(\alpha_k$) for $1\leq k
\leq \infty$, i.e. the implementation of $U(\alpha)$ with arbitrary $\alpha$
($0\leq\alpha\leq \pi$). Using (\ref{Sl}), one finds that the total number of
qubits needed to store one of those operations is on average given by
\be
\sum_{k=1}^{\infty} S_k \sum_{l=0}^{k-1} \frac{1}{2^{l}} \leq 2 \sum_{k=1}^{\infty} S_k,\label{Sk}
\ee
which can evaluated to be 3.8942. That is, less than 4 qubits per operation are
required on average to store an arbitrary, unknown operations of the form
$U(\alpha)$. In \cite{Vi00}, it was found that on average two qubits suffice to
store $U(\alpha)$.

However, if we restrict the possible values of $\alpha$ to $0\leq\alpha\leq
\pi/8$ [$\pi/32$], we find that the average amount of required storage qubits is
given by 0.2257 [0.0206]. This can be seen by noting that in this case, the sum
in Eq. (\ref{Sk}) starts with $k=4$ [$k=6$] respectively. Thus we showed that
unitary operations of the form $U(\alpha)$ can be stored locally, and that the
average amount of qubits required for storage can be decreased if one restricts
the operations to be stored. This result is similar to the one obtained by
Schumacher \cite{Sc97} for the compression of a set of pure states.

\subsubsection{Compression of a finite set of unitary operations}\label{fin}

Here, we consider a finite set of unitary operations of the form
$U(\alpha_N)$ (\ref{Ualpha}), where $\alpha_N=\pi/2^N$ and $1\leq
N\leq M$ and provide an alternative protocol for storage and
compression. This set of operations can be viewed as the basic set
required to implement arbitrary operations. We assume that each of
the operation is equally likely. Again, we follow the procedure
described in Sec. \ref{NLU} {\bf (i-ii)}, in order to implement a
certain operation of the form (\ref{Ualpha}), say $U(\alpha_N)$,
with unit probability. The set of states $G_N$ (\ref{S_N}) is
required, where the corresponding probabilities are given by
$p_k=1/2^{k-1}$ for the $k^{\rm th}$ state. Note that for
different $N$, a different number of steps are required and thus a
different number of states has to be stored. As this may cause
problems, we fix the number of states to be stored for each
operation to be $M$. In case less than $M$ steps are required, the
state $|\Psi_0\rangle$ is stored in the remaining cases, which
corresponds to the identity operation. Now, the implementation of
any operation $U(\alpha_N)$ consists of at most $M$ steps, where
in steps $(N+1),\ldots,M$ the identity operations is performed.
The following table summarizes the states which are stored for
each of the operations:
\bea
U(\alpha_1): &&G_1=\{|\Psi_{\frac{\pi}{2}}\rangle,|\Psi_{0}\rangle,|\Psi_{0}\rangle,|\Psi_{0}\rangle,\ldots,|\Psi_{0}\rangle\}\nonumber\\
U(\alpha_2): &&G_2=\{|\Psi_{\frac{\pi}{4}}\rangle,|\Psi_{\frac{\pi}{2}}\rangle,|\Psi_{0}\rangle,|\Psi_{0}\rangle,\ldots,|\Psi_{0}\rangle\} \nonumber \\
U(\alpha_3): && G_3=\{|\Psi_{\frac{\pi}{8}}\rangle,|\Psi_{\frac{\pi}{4}}\rangle,|\Psi_{\frac{\pi}{2}}\rangle,|\Psi_{0}\rangle,\ldots,|\Psi_{0}\rangle\}\\
\ldots&&\nonumber\\
U(\alpha_M): &&G_M=\{|\Psi_{\alpha_M}\rangle,|\Psi_{2\alpha_M}\rangle,|\Psi_{4\alpha_M}\rangle,\ldots,|\Psi_{\frac{\pi}{2}}\rangle\}\nonumber.
\eea
Recall that the $k^{\rm th}$ state is always used in the $k^{\rm
th}$ step. We denote by $C_k$ the $k^{\rm th}$ column, which
consists of the $k^{\rm th}$ element of each $G_l$. As the
columns $C_k$ correspond to the different steps, we have that
column $k$ is only required with probability $p=1/2^{k-1}$ and
all steps ---and thus all columns $C_k$--- can be treated
independently. That is, we store each of the columns $C_k$
independently. Due to the fact that all states within each
column $C_k$ are equal likely and non--orthogonal, one can use
data compression \cite{Sc97}. The compression factor $S_k$ for
column $C_k$ is given by the entropy of the density operator
$\rho_k$, where $\rho_k$ is an equal mixture of all the states
of column $k$, and column $k$ is only required with probability
$p=1/2^{k-1}$. Thus the total number of qubits required to store
one of the operations $U(\alpha_N)$, $0\leq N\leq M$ is given by
$\sum_{k=1}^{M}S_k/2^{k-1}$. For example, for $M$=100 [1000] we
obtain an average amount of 0.245 [0.0361] qubits which have to
be stored on average to implement one of the 100 [1000]
operations picked at random.

%-------------------------------------------------------------------------
%-------------------------------------------------------------------------
\subsection{Storage of non--local unitary operations in $A$ and $B$}\label{nonloc}

Here, we consider {\bf (ii)}, the storage of a set of non--local unitary
operations. We will discuss variations of both protocols described in the
previous section, taking into account that we now have two spatially separated
parties and the operations are non--local. That is, the states to be stored are
entangled states and we consider local storage of the subsystem belonging to $A$
($B$). This means that both, the coding and decoding procedure has to be local,
but may be assisted by classical communication.

We first consider the storage of an infinite set of unitary operations of the
form $U(\alpha)$ (see Sec. \ref{inf}). We follow the same protocol as described
in Sec. \ref{inf}, however we now use a different kind of data compression. The
protocol described in Sec. \ref{inf} involves storage of two equal likely
states, $|\Psi_{0}\rangle$ or $|\Psi_{\alpha_N}\rangle$ for some
$\alpha_N=\pi/2^N$. Note that the state $|\Psi_{\alpha_N}\rangle$ is an
entangled state, so in contrast to the previous section, we cannot use standard
data compression for pure states, as we are restricted to local operations only.
However ---as shown in Appendix A--- it is also possible to achieve {\it local}
data compression for a set of entangled states. That is, each of the parties
manipulates only its own subsystem and can thereby reduce the average number of
qubits required to store its part of the entangled state, without affecting the
entanglement with the other system. Note that this problem is equivalent to the
data compression of mixed states with commuting density operators, where the
entanglement with some other system should be preserved. It turns out that the
compression factor for $A$ ($B$) is given by the entropy of an operator $\tilde
\rho$, which is an equal mixture of the reduced density operators $\rho^i_A$
($\rho^i_B$) corresponding to the states $|\Psi_{0}\rangle$,
$|\Psi_{\alpha_N}\rangle$. Note that this corresponds to the
upper bound on the number of qubits to be stored in case the
entanglement with another system is not required to be preserved
\cite{Ba00}. While it is known that this is not the optimal
compression rate in case the entanglement with some other system
is not required to be preserved, it is not clear whether the
compression rate is already optimal under this stronger
restriction. In our specific case, we obtain
\be
S_N=-x_N \log_2(x_N)-(1-x_N)\log_2(1-x_N),
\ee
with $x_N=(1+\cos^2{\alpha_N})/2$. Using now this local
compression protocol instead of Schumacher's for pure states in
the protocol of Sec. \ref{inf}, one finds that the the average
number of qubits which have to be stored locally in $A$ ($B$) is
given by 4.7758 if $0\leq\alpha\leq\pi$. If we restrict the
possible values of $\alpha$ to $0\leq\alpha\leq\pi/8$
[$\pi/32$], we find that the average amount of required storage
qubits is reduced to 0.3976 [0.0379].

Regarding the storage of a finite set of unitary operations (see Sec.
\ref{fin}), we follow the same protocol as described in Sec. \ref{fin} and use
again a different kind of data compression due to the fact that we are
restricted to local operations. This time, data compression for a finite set
$\{|\Psi_i\rangle\}$ of $M$ entangled states is required. The entangled states
are all equal likely and are of the form $|\Psi_{\alpha_N}\rangle$. It turns out
(see Appendix A) that one can achieve a compression rate which is given by the
entropy of a density operator $\tilde \rho$, which is defined as an equal
mixture of the reduced density operators $\rho_A^i$ corresponding to the state
$|\Psi_i\rangle$. One finds that the total number of qubits required on average
to store one of the operations $U(\alpha_N)$, $0\leq N\leq M$ locally in $A$
($B$) is given by 0.333 [0.050] qubits for $M$=100 [1000].

%-------------------------------------------------------------------------
%-------------------------------------------------------------------------
\subsection{Storage of non--local unitary operations in $A$}

Finally, we consider {\bf (iii)} the local storage of a non--local unitary
operation in $A$. That is, we consider a local memory (in $A$ only), but we want
to implement the operation non--locally. It turns out that this problem is a
trivial combination of the previous two problems. We have that one can use the
methods described in Sec. \ref{loc} to store the operations locally in $A$, and
one obtains the the same compression rates. The average amount of quantum
communication from $A$ to $B$ ---which is required to implement the operation
non--locally--- can be found using the method described in Sec. \ref{nonloc}.
That is, one part of the entangled system is compressed and send through a
quantum channel to $B$. The compression rate can be calculated in a similar way
as in Sec. \ref{nonloc}, however one has to take into account that the state
$|\Psi_{\pi/2}\rangle$ is a separable state and thus no quantum communication is
required to transmit one part of this state. For example, one finds in case of
an infinite set of operations of the form $U(\alpha)$ with $0\leq \alpha\leq
\pi$ [$\pi/8$] that the required amount of quantum communication from $A$ to $B$
is given by 2.7758 [0.3976] qubits.

This last method clearly distinguishes between the required amount of local
storage qubits and the non--local content of the operation, i.e. the average
amount of quantum communication. Note that storing the operations locally (see
Sec. \ref{loc}) requires a smaller amount of storage qubits than storing a
non--local operation directly in $A$ and $B$ (see Sec. \ref{nonloc}).

%-------------------------------------------------------------------------
%-------------------------------------------------------------------------
\section{Non--local Measurements}\label{NLP}
%-------------------------------------------------------------------------
%-------------------------------------------------------------------------

In this Section, we consider the implementation of non--local
measurements. We consider two spatially separated parties, $A$
and $B$, each possessing a $d$--level system. The two parties
want to perform a complete, joint measurement on their system,
specified by a set of rank $n_k$ Projectors $\{P_k\}$ such that
$\sum_k P_k^{AB} =\eins_A\otimes\eins_B$. The questions we pose
are the following: How can the parties implement this non--local
measurement? What are the entanglement properties of those
measurements; that is, (i) what is the amount of entanglement
required to implement a certain measurement? (ii) what is the
average amount of entanglement which can be produced given a
single application of the non--local measurement?

We provide several procedures to implement arbitrary non--local von Neumann
measurements and discuss their entanglement properties. We show that the
required amount of entanglement depends on the measurement to be implemented. We
introduce examples of non--local measurements which can be implemented using
less than one ebit of entanglement. One can easily generalize some of our
results to implement also arbitrary measurement, described by a positive
operator valued measure (POVM), i.e. a set of positive operators $O_k^{AB}$ such
that $\sum_k O_k^{AB} =\eins_A\otimes\eins_B$.

First, we note that the amount of entanglement required to implement the
non--local measurement depends on {\bf (i)} whether one is only interested in
the measurement outcome or {\bf (ii)} the system should in addition be in the
corresponding state after the measurement. For example, one can perform a
complete Bell measurement (i.e. a measurement in the basis (\ref{Bell})) on a
state of two qubits using one ebit of entanglement regarding {\bf (i)}, while
two ebits are required in case of {\bf (ii)}.

{\bf Proposal 1:}
A trivial procedure to perform an arbitrary bipartite measurement is the
following: The state of system $B$ is teleported to $A$, consuming $\log_2(d)$
ebits. Then, the measurement is performed locally in $A$, which already suffices
in case of {\bf (i)}. Regarding {\bf (ii)}, one also has to teleport the
particle back to $B$, consuming again $\log_2(d)$ ebits. Note that in case of a
complete Bell measurement, i.e. a measurement in the basis (\ref{Bell}), where
each basis state is a MES, this procedure is in fact optimal. On the one hand, one
consumes two ebits to implement the measurement. On the other hand, one can also
obtain an average amount of two ebits given a single application of a non--local
Bell measurement. One just has to consider the operator $E_i$
(\ref{iso}) associated to each possible outcome of the Bell measurement. One
observes that the non--local entanglement of all $E_i$ is given by two ebits,
and each measurement outcome is equal likely. This leads to an average amount of
entanglement of two ebits. We have that if the amount of entanglement required
to implement an operation, $E_{\rm im}$ equals the amount of entanglement which
can be obtained given a single application of the operation, $E_{\rm cr}$, the
first process is optimal and $E_{\rm im}$ is the minimal amount of entanglement
required to implement the operation. This is due to the fact that $E_{\rm cr}
\leq E_{\rm im}$, otherwise one could create entanglement for free.
However, if one wants to measure the joint system $A$,$B$ in a basis which is
not maximally entangled, one might expect that the required amount of
entanglement is smaller than $2\log_2(d)$ ebits. With the following method, we
show that this is indeed the case.

{\bf Proposal 2:}
We consider the situation were all $P_k$ are
rank one, i.e. $n_k=1$ and thus $P_k=|\phi_k\rangle_{AB}\langle\phi_k|$. We define a
non--local unitary operation $U$ by
\be
U=\sum_{k=1}^{d^2} |k\rangle_{AB} \langle \phi_k|,
\ee
where $|k\rangle_{AB}=|a_k\rangle_A|b_k\rangle_B$, and $\{|a_i\rangle\}$
[$\{|b_i\rangle\}$] is some local basis in $A$ [$B$] respectively, with $1\leq k
\leq d$.  The procedure takes place as follows: First, the parties apply the
non--local unitary operation $U$, using e.g. the procedure described in Sec.
\ref{NLU} for $d=2$, consuming an amount of entanglement which is specified by
the operation $U$. In case $U$ is only weakly entangling, e.g. if $\langle
k|\phi_k\rangle \approx 1$ (i.e. $|\phi_k\rangle$ are only weakly entangled
states), the required entanglement is small (see Sec. \ref{NLU}) \cite{example}.
Then, parties $A$ and $B$ both perform a local measurement specified by
projectors on the states $\{|a_i\rangle\}$ [$\{|b_i\rangle\}$] respectively and
communicate the outcome of the measurement classically. In case they obtained
the outcome $a_k,b_k$, they know that the outcome of the measurement is $k$,
i.e. they measured the projector $P_k$. Concerning {\bf (i)}, the procedure ends
at this point. Regarding {\bf (ii)}, $A$ and $B$ in addition implement the
operation $U^\dagger$ to ensure that the system is also in the required state
after the measurement. Alternatively, they could also prepare the measured
system in state $|\phi_k\rangle$, as due to the implementation of the
measurement, any possible entanglement with some auxiliary system is destroyed
anyway. We note that the choice of the local basis in $A$, $\{|a_i\rangle\}$ and
$B$, $\{|b_i\rangle\}$ is not fixed and may also change the entanglement
properties of the operation $U$. This can be seen by considering the following
trivial example: We have $d=2$ and $|\phi_{00}\rangle=|00\rangle,
|\phi_{01}\rangle=|01\rangle, |\phi_{10}\rangle=|10\rangle$ and
$|\phi_{11}\rangle=|11\rangle$. By choosing $|a_1\rangle |b_1\rangle =|0\rangle$
and $|a_2\rangle |b_2\rangle =|1\rangle$, we have that $U = \eins_{AB}$, i.e. no
entanglement is required to perform the measurement. If we however choose the
mapping $|\phi_{00}\rangle \rightarrow |00\rangle, |\phi_{01}\rangle \rightarrow
|10\rangle, |\phi_{10}\rangle \rightarrow |01\rangle, |\phi_{11}\rangle
\rightarrow |11\rangle$, we find that the operation $U=U_{\rm swap}$, which
requires two ebits to implement \cite{Co00}. In this case, the choice of the
proper local basis is trivial, however we do not know the optimal choice for a
general measurement. Note also that this procedure fails to implement non--local
measurements where the rank of some projector $P_k$ is larger than one. For
example, if $P_1=|00\rangle_{AB}\langle 00|$ and $P_2=\eins_{AB}-P_1$, this
procedure fails to project in the subspace spanned by $P_2$, as it already gives
a fine--graining within this subspace, which is a different problem. The
next method will overcome this limitation.

{\bf Proposal 3:} Here, we consider a complete set of $M$
non--local projectors $P_k$ which might have arbitrary rank $n_k$.
Clearly, $\sum_{k=1}^{M}n_k=d^2$. Alice uses an $M$--level
auxiliary system initially prepared in state $|1\rangle$, which is
used to label all possible measurement outcomes. We define a
unitary operation $U$ acting on the auxiliary system $\tilde A$
and the joint system $AB$ as follows:
\bea
U&=&\sum_{j=1}^M [(|j\rangle_{\tilde A}\langle 1|+ |1\rangle_{\tilde A}\langle j|)\otimes P_j^{AB} \nonumber\\
&& +(\eins_{\tilde A} - |1\rangle_{\tilde A}\langle 1| - |j\rangle_{\tilde A}\langle j|)\otimes P_j^{AB} ].
\eea
 %The relevant part is given by the first term, which states that system $\tilde
 %A$ is in state $|j\rangle$ if the projector $P_j$ is applied to system $AB$.
After application of $U$, the auxiliary system $\tilde A$ is measured in
the basis $\{|j\rangle\}$. If the outcome $k$ is found, one readily observes
that this corresponds to measuring the projector $P_k$ on the system $AB$. Note
that no further operations are required, as the system $AB$ is already in the
appropriate state {\bf (ii)}. The amount of entanglement required to implement
the non--local measurement is again specified by the operation $U$.

For example, if $d=2$ and $P_1=(|00\rangle\langle00|+|11\rangle\langle11|)$,
$P_2=(|01\rangle\langle 01|+|10\rangle\langle10|)$, it turns out that one can
create one ebit given a single measurement of this kind. To see this, we prepare
the system $AB$ in the product state
$\rho=1/2(|\Phi^+\rangle\langle\Phi^+|+|\Psi^+\rangle\langle\Psi^+|)$ and perform
the measurement. In case we obtain outcome ``1'' [``2''], the state after the
measurement is $|\Phi^+\rangle$ [$|\Psi^+\rangle$] respectively. In both cases,
we created one ebit. However, it is not clear whether one ebit of entanglement
also suffices to implement the corresponding unitary operations $U=\eins^{\tilde
A}\otimes P_1^{AB}+ \sigma_x^{\tilde A}\otimes P_2^{AB}$. Although the state
$E_U$ associated to $U$ via (\ref{iso}) has an amount of entanglement of one
ebit and $U=U^\dagger$, it is not clear whether a single copy of the state $E_U$
suffices to implement $U$.

It would be interesting to establish the minimal amount of entanglement required
to implement a general, non--local measurement.

%-------------------------------------------------------------------------
%-------------------------------------------------------------------------
\section{Multiparty operations}\label{multi}
%-------------------------------------------------------------------------
%-------------------------------------------------------------------------

In this Section, we generalize some of the previous results to
multiparty systems. We consider several spatially separated
systems $A,B, \ldots ,Z$, each possessing several $d$--level
systems. We first generalize the isomorphism (\ref{iso}) between
CPM ${\cal E}$ and positive operators $E$ to multiparty systems.
Here, ${\cal E}$ acts on several $d$--level systems, one located
in each site $A, B,\ldots ,Z$ and $E$ is a positive operator on
the Hilbert space ${\cal H}_{A_{1,2}}\otimes \ldots \otimes{\cal
H}_{Z_{1,2}}$. We have that ${\cal H}_{A_i}=\C^d$ and similar for
the remaining parties. For a $N$--party system, it is easy to show
that
\bma\label{isoN}\bea
E_{A_{1,2}\ldots Z_{1,2}} &=& {\cal E}(P_{A_{1,2}}\otimes \ldots \otimes P_{Z_{1,2}}),\label{iso1N}\\
{\cal E}(\rho_{A_1\ldots Z_1})&=& d^{2N}{\rm tr}_{A_{2,3}\ldots
Z_{2,3}} \nonumber\\&& (E_{A_{1,2}\ldots Z_{1,2}} \rho_{A_3\ldots
Z_3} P_{A_{2,3}}\ldots P_{Z_{2,3}})\label{iso2N}.
\eea\ema
The interpretation is similar to the one of Eq. (\ref{iso}). On one hand,
(\ref{iso1N}) states that $E$ can be created from a $N$--party product state,
where each party prepares locally a MES. On the other hand, (\ref{iso2N}) tells
us that given $E$ (particles $A_{1,2}B_{1,2}\ldots Z_{1,2})$, one can implement
the multi---particle operation ${\cal E}$ on an arbitrary state $\rho$ of $N$
$d$--level systems (particles $A_3B_3\ldots Z_3$) by measuring locally the
projector $P$ (\ref{P}) on particles $A_{2,3},B_{2,3},\ldots ,Z_{2,3}$ in each
of the locations. Note that the probability of success is given by $p=1/d^{2N}$.

As in the bipartite case, one may ask for a certain map ${\cal E}$ whether it is
capable to create entanglement. Since for multiparty systems, there exist many
different kinds of entanglement (see e.g. \cite{Du00W,Du00A,Du00BE,rest}), one
may also ask which kind of entanglement can be produced. Again, all these questions
can be answered by establishing the entanglement properties of the operator $E$
associated to the CPM ${\cal E}$ via (\ref{iso1N}). In particular, if $E$ is
bound entangled \cite{Du00BE}, then ${\cal E}$ can only create BES.
In a similar way, given some BES one can easily construct the corresponding map
which is capable of generating BES of the same kind.

One may also consider the implementation of arbitrary $N$--qubit
operations with unit probability. On one hand, any $N$--qubit
operation can be written as a sequence of bipartite CNOT
operations and single qubit unitary operations, for which we
already established a protocol. On the other hand, we may consider
$N$--qubit unitary operations of a specific form and show directly
how to implement them with unit probability given certain states.
We consider a unitary operation of the form
\be
U_N(\alpha_M)=e^{-i\alpha_M \sigma_x^{A_1}\otimes \ldots \otimes
\sigma_x^{Z_1}}\label{NUalpha},
\ee
where $\alpha_M=\pi/2^M$. It turns out that a natural extension of the protocol
of Sec. \ref{NLU} {\bf (i-iii)} allows to implement operations of the form
(\ref{NUalpha}) with probability $p=1$. The operator associated with the unitary
operation $U_N(\alpha_M)$ is given by $E_{A_{1,2},\ldots,Z_{1,2}}=
|\psi_{\alpha_M}\rangle \langle \psi_{\alpha_M}|$, where
\bea\label{psialphaN}
|\psi_{\alpha_M}\rangle &=& \cos(\alpha_M)|\Phi^+\rangle_{A_{1,2}}
|\Phi^+\rangle_{B_{1,2}}\ldots|\Phi^+\rangle_{Z_{1,2}}\nonumber\\&& -
i\sin(\alpha_N)|\Psi^+\rangle_{A_{1,2}} |\Psi^+\rangle_{B_{1,2}}\ldots|\Psi^+\rangle_{Z_{1,2}}.
\eea
Regarding {\bf (i)}, we just note that Bell measurements and the corresponding
local unitary operations are performed at all location $A,B,\ldots,Z$. For all
possible measurement outcomes, it is easy to observe that the operation
performed on some state $\rho_{A_1,\dots,Z_1}$ will either be (i) $U(\alpha_M)$
or (ii) $U(-\alpha_M)$, each possibility appearing with probability $p=1/2$.
Steps {\bf (ii)} and {\bf (iii)} can be adopted without changes, which finally
allows to implement an operation of the form (\ref{NUalpha}) with arbitrary
angle $\alpha$ and unit probability. Note that operations (\ref{NUalpha}) are
capable to create GHZ--like entanglement and are thus truly $N$--qubit
entangling operations.

%-------------------------------------------------------------------------
%-------------------------------------------------------------------------
\section{Summary}\label{summary}
%-------------------------------------------------------------------------
%-------------------------------------------------------------------------

To summarize, we have provided several applications of an previously introduced
isomorphism between operations an states. First, we discussed how to use this
isomorphism to establish separability and entangling properties of operations
${\cal E}$ and to construct physical operations which are capable to create
bound entangled states. In addition, we showed how to implement an arbitrary
non--local two--qubit operation consuming an amount of entanglement which is
proportional to the entangling capability of the operation.

Then, we have shown how to implement several techniques
developed for states ---such as purification or data
compression--- also for operations. In particular, we have shown
that a known, noisy, non--local unitary operation as well as an
unknown, noisy, local unitary operation can be purified. In a
similar way, we use these results to establish tomography of
arbitrary operations. Then, we showed that unitary operations
can be stored locally and non--locally and that the amount of
required qubits for storage can be decreased, which can be
viewed as generalization of data compression to unitary
operations. In this context, we also provided a protocol which
allows for local data compression of a set of entangled states.
Note that it is straightforward to obtain a number of other
results which were developed for states also for operations. For
example, it is easy to show that also unitary operations can be
cloned (via cloning of the corresponding state $E$) or
teleported (via teleportation of the states required to store
the operation) \cite{Hu00,Vi00}. In case of cloning, one has to
take into account that the cloned states allow for a
probabilistic, imperfect implementation of the required
operation only.

We also provided a method to implement arbitrary two--photon gates
probabilistically with present day technology, which opens the way
for practical quantum communication over arbitrary distances.
Finally, we discussed the implementation of non--local
measurements and generalized some of our results to multi--party
systems.

%-------------------------------------------------------------------------
\section*{Acknowledgements}
We thank B. Kraus, G. Giedke and G. Vidal for interesting discussions. This work
was supported by the Austrian SF under the SFB ``control and measurement of
coherent quantum systems'' (Project 11), the European Community under the TMR
network ERB--FMRX--CT96--0087 and project EQUIP (contract IST-1999-11053), the
ESF, and the Institute for Quantum Information GmbH.

%-------------------------------------------------------------------------
%-------------------------------------------------------------------------
\section*{Appendix A: Local data compression for a set of entangled states}
%-------------------------------------------------------------------------
%-------------------------------------------------------------------------

In this appendix, we consider the problem of local data compression of a set of
pure, entangled states, where all reduced density operators commute. Note that
this problem is equivalent to the problem of data compression of a set of
commuting mixed states under the restriction that entanglement with some other
systems should be preserved. Let $G=\{|\Psi_i\rangle\}_{i=1}^{L}$ be a set of $L$
pure states, where
\be
|\Psi_i\rangle=c_{\alpha_i}|00\rangle_{AB}+s_{\alpha_i}|11\rangle_{AB}.
\ee
and $c_{\alpha_i}\equiv \cos(\alpha_i)$, $s_{\alpha_i} \equiv \sin(\alpha_i)$.
Each state is assigned a prior probability $p_i$. Two spatially separated
parties $A$ and $B$ are fed an unending sequence of states $|\Psi_j\rangle$,
where each successive state is chosen randomly and independently from the set
$G$ according to the probability distribution $\{p_i\}$. A sequence of length $N$ is of the form
\be
|\Psi_{i_1i_2\ldots i_N}\rangle=|\Psi_{i_1}\rangle|\Psi_{i_2}\rangle\ldots |\Psi_{i_N}\rangle,\label{sequ}
\ee
and appears with probability $p_{i_1i_2\ldots i_N}=p_{i_1}p_{i_2}\ldots
p_{i_N}$. The parties $A$ and $B$ want to store the sequences locally, i.e they
are allowed to perform local operations and classical communication. We are
interested in the average amount of qubits per signal state which are required
in $A$ ($B$) to store the signals faithfully. We will use as a criterium the
so--called GLOBAL-FID criterium \cite{Ba00}; that is we require that the average
global fidelity of all possible sequences is $1-\epsilon$. Note that we consider
the so called ``blind case'' \cite{Ba00}, that is neither $A$ nor $B$ know the
specific sequence (\ref{sequ}).

Let $\rho_i^A={\rm tr}_B (|\Psi_i\rangle\langle\Psi_i|)$ be the reduced density
operator of system $A$ of the state $|\Psi_i\rangle$ and
\be
\tilde\rho^A=\sum_{i=1}^{L} p_i\rho_i^A\label{trho}
\ee
be the weighted average of the reduced density operators of our signal source.
We denote by $S(\tilde\rho^A)={\rm tr}(\tilde\rho^A \log_2\tilde\rho^A)$ be the von Neumann
entropy of $\tilde \rho^A$.

Given a sequence of length $N$, $N$ sufficiently large, we provide a protocol with the following properties:
\begin{description}
\item[{\bf (i)}] The required amount of storage qubits in $A$ ($B$) is given by $N S(\tilde\rho^A)+\delta$.
\item[{\bf (ii)}] The average global fidelity (averaged over all possible sequences) $\bar F$ is given by $1-\epsilon$.
\end{description}
We have that $\delta$ is some function which is of the form $\delta = \mu 
N^\beta$ for some  $\mu>1, 1/2<\beta<1$ and $\epsilon\rightarrow 0$ as 
$N\rightarrow\infty$. That is, on average $S(\tilde\rho^A)$ qubits per signal 
state have to be stored locally in $A$ ($B$). Note that we do not claim that 
this the optimal compression rate achievable.

For pedagogical reasons, we will proof our statement in the simplest case,
where the set $G$ consists of two pure states only. We will even assume that
$|\Psi_1\rangle=|00\rangle$ and
$|\Psi_2\rangle=c_{\alpha}|00\rangle+s_{\alpha}|11\rangle$ and $p_1=p_2=1/2$.
Note that the proof can be easily generalized to an arbitrary number of signal
states and an arbitrary probability distribution.

We have that
\be
\tilde\rho^A=\frac{1+c_\alpha^2}{2}|0\rangle\langle 0|+\frac{1-c_{\alpha}^2}{2}|1\rangle\langle 1|.
\ee
We define local projectors $P_A$ ($P_B$) acting on $N$ qubits as follows:
\be
P_A=P_B=\sum_{k=k^-}^{k^+} P_k,
\ee
where $k^\pm=(1+c_\alpha^2)/2 \pm \mu N^\beta$, $\mu>0$, $1/2<\beta<1$ and
\be
P_k=\sum_{\rm perm} |0\rangle\langle 0|^{\otimes k} \otimes|1\rangle\langle 1|^{\otimes N-k}.\label{Pk}
\ee
The sum in (\ref{Pk}) runs over all possible $b_{N,k}\equiv N!/[k!(N-k)!]$
permutations (without repetitions) of $k$ zeros and $N-k$ ones. Thus $P_k$ is a
projector in the subspace spanned by all states which contain exactly $k$ zeros
and $(N-k)$ ones. The dimension of $P_k$ is given by $b_{N,k}$.

The projector $P_A$ ($P_B$) is measured locally in $A$ ($B$). If the measurement
is successful, $\log_2 (d)$ ---where $d={\rm dim}(P_A))$--- qubits are used to
store the resulting state in $A$. This can be accomplished by relabeling the
states which span $P_A$ to $\{|l\rangle\}_{l=1}^{d}$ and store those states locally,
which clearly requires $\log_2(d)$ qubits. The decoding procedure consists of
undoing the relabeling. In case the measurement is not successful, some state $|0_E\rangle$ is
stored instead. We show that: {\bf (i)} $\log_2(d) = NS(\tilde
\rho) +\delta$ and {\bf (ii)} $\bar F= \sum _{i_1i_2 \ldots i_N} p_{i_1i_2
\ldots i_N} F_{i_1i_2\ldots i_N} > 1-\epsilon$, where $F_{i_1i_2\ldots
i_N}=|\langle\Psi_{i_1i_2\ldots i_N}|P_A\otimes P_B|\Psi_{i_1i_2\ldots
i_N}\rangle|^2$ and the sum runs over all possible sequences.
That is, the storage procedure requires the announced amount of qubits and the
average fidelity is sufficiently large.

Regarding {\bf (i)}, it is easy to see that $d={\rm dim}(P_A) \leq
(k^-+k^++1)b_{N,k_0}$, where $k_0\in[k^-,k^+]$ is the value that maximizes
$b_{N,k}$ in this interval. Substituting the values of $k^-,k^+$ in this bound,
we find that $\log_2(d) = N S(\tilde \rho^A) + \delta$ as required.

We now concentrate on {\bf (ii)}, the average global fidelity $\bar F$. Consider
a sequence of the form (\ref{sequ}) which contains $j$ states $|\Psi_1\rangle$
and $(N-j)$ states $|\Psi_2\rangle$ (i.e. the number of $i_k$ which are equal to
one is given by $j$). We denote such a sequence by $|\Psi(j)\rangle$. Note that
there are $b_{N,j}$ sequences of this kind. For all those sequences, we find
\bea
F_j&=&|\langle\Psi(j)|P_A\otimes P_B|\Psi(j)\rangle|^2 \nonumber\\
&=&|\sum_{k; k^-\leq(j+k)\leq k^+}c_\alpha^{2k}s_\alpha^{2(N-k)} b_{N-j,k}|^2,\label{Fj}
\eea
The average fidelity $\bar F$ is given by
\be
\bar F=\frac{1}{2^N} \sum_{k^-\leq j\leq k^+} b_{N,j} F_j,\label{boundF}
\ee
where $2^{-N}$ is the probability that a certain sequence appears, $b_{N,j}$ is the
number of sequences of the form $|\Psi(j)\rangle$ and $F_j$ is given in (\ref{Fj}).
We have that
\be
\bar F \geq \sum_{j=j^-}^{j^+} \frac{1}{2^j} \frac{1}{2^{N-j}} b_{N,j} F_j,\label{newbarF}
\ee
where $j^\pm=N/2\pm \mu/3 N^\beta$. In this case, one can also bound $F_j$ and finds
\be
F_j \geq |\sum_{k=\tilde k^-}^{\tilde k^+}c_\alpha^{2k}s_\alpha^{2(N-k)} b_{N-j,k}|^2,
\ee
where $\tilde k^\pm=c_\alpha^2(N-j)\pm\mu/3 N^\beta$ and we have that $k^-\leq (j+k)
\leq k^+$ as required. By noting that a binomial distribution is asymptotically
equivalent to a normal (Gaussian) distribution, the fidelity $F_j$ $\forall j$
can be seen to be bounded from below by $\Phi(2\mu N^{\beta-1/2})$, where
$\Phi(x)\equiv1/\sqrt{2 \pi} \int_{-x}^x e^{y^2/2}dy$. For our choice of $\mu,
\beta$, we have that $F_j \rightarrow 1$ when $N \rightarrow \infty$. In a
similar way, one shows that also $\bar F \rightarrow 1$ when $N \rightarrow
\infty$, as after bounding $F_j$ as stated above, (\ref{newbarF}) also
corresponds to a binomial distribution centered at $j=N/2$. This finishes the
proof of the statements {\bf (i-ii)}.

In a similar way, one can carry out the analysis for a set of $L$ entangled
states and an arbitrary probability distribution $\{p_i\}$. In this case,
$k^\pm=N \sum_{i=1}^{L} p_i c_{\alpha_i}^2 \pm \mu N^\beta$ and some of the
binomial distributions are replaced by multinomial distributions. Also in this
case, one finds that $\bar F \rightarrow 1$ for $N \rightarrow \infty$ and that
the dimension of the projector $P_A$ ($P_B$) is given by the $N$ times the entropy
of the operator $\tilde \rho^A$ (\ref{trho}).

%-------------------------------------------------------------------------
%-------------------------------------------------------------------------


\begin{references}
\bibitem{note1}
If $\rho$ is a multiqubit state and $U$ is performed on qubits one and two, the
noisy operation is defined as follows:  ${\cal E}(\rho) = q U_{12}\rho
U_{12}^\dagger + \frac{1-q}{16}\text{tr}_{12}\{\rho\} \otimes  \eins_{12},$ i.e.
the state of the remaining particles is described by their reduced density
operator. This noise model was also used in the context of quantum repeaters \cite{Br98}.
\bibitem{noteCNOT}
The CNOT operation is defined by the following mapping of states, written in the
standard basis: $|i\rangle_A|j\rangle_B \rightarrow |i\rangle_A|i\oplus
j\rangle_B$, where $\oplus$ denotes addition modulo 2.
\bibitem{note2}
The 4--qubit state $\rho_{\cal E}$ can in an appropriate basis be
written as $q|\psi^-\rangle_{AB}\langle\psi^-|+(1-q)/16\eins_4$.
One can easily check that the partial transposition of $\rho_{\cal
E}$ is non positive iff $q>1/9$. Projecting the state $\rho_{\cal
E}$ locally ---both in a $A$ and $B$--- in the subspace spanned by
$\{|0\rangle,|1\rangle\}$, the resulting state is of Werner form
\cite{We89} and can be distilled to singlet form if $q>1/9$
\cite{Be96}. Due to the fact that non positive partial
transposition is a necessary condition for distillability, we have
that $\rho_{\cal E}$ is distillable to singlet form if and only if
$q>1/9$.
\bibitem{ortho} We are considering Hilbert space of operators with scalar
product $(A_i,A_k)$=tr$(A_i^\dagger A_k)$. Two operators $A_i$ and $A_k$ are orthonormal if
$(A_i,A_k)=\delta_{i,k}$.
\end{references}
\end{document}